# A high-performance virtual machine filesystem monitor in cloud-assisted cognitive IoT

Dongyang Zhan[a], Lin Ye[a], Hongli Zhang[a], Binxing Fang[a,b], Huhua Li[a], Yang Liu[a], Xiaojiang Du[c], Mohsen Guizani[d]

[a]*School of Computer Science and Technology, Harbin Institute of Technology, Harbin, 150001, China*
[b]*Institute of Electronic and Information Engineering of UESTC in GuangdongDongguan Guangdong 523808, China*
[c]*Department of Computer and Information Sciences, Temple University, Philadelphia, PA 19122, USA*
[d]*Department of Electrical and Computer Engineering, University of Idaho, Moscow, ID, USA*

**Abstract**

Cloud-assisted Cognitive Internet of Things has powerful data analytics abilities based on the computing and data storage capabilities of cloud virtual machines, which makes protecting virtual machine filesystem very important for the whole system security. Agentless periodic filesystem monitors are optimal solutions to protect cloud virtual machines because of the secure and low-overhead features. However, most of the periodic monitors usually scan all of the virtual machine filesystem or protected files in every scanning poll, so lots of secure files are scanned again and again even though they are not corrupted. In this paper, we propose a novel agentless periodic filesystem monitor framework for virtual machines with different image formats to improve the performance of agentless periodic monitors. Our core idea is to minimize the scope of the scanning files in both file integrity checking and virus detection. In our monitor, if a file is considered secure, it will not be scanned when it has not been modified. Since our monitor only scans the newly created and modified files, it can check less files than other filesystem monitors. To that end, we propose two monitor methods for different types of virtual machine disks to reduce the number of scanning files. For virtual machine with single disk image, we hook the backend driver to capture the disk modification information. For virtual machine with multiple copy-onwrite images, we leverage the copy-on-write feature of QCOW2 images to achieve the disk modification analysis. In addition, our system can restore and remove the corrupted files. The experimental results show that our system is effective for both Windows and Linux virtual machines with different image formats and can reduce the number of scanning files and scanning time.

**1. Introduction**

Assisted by cloud computing, Cognitive Internet of Things (CIoT) has some new capabilities such as machine learning and data mining. Since the huge amount of data generated by IoT is stored and processed by virtual machines (VMs) in cloud, the VM security, especially VM filesystem security, is very important for the whole CIoT system. Attackers usually tamper with critical files (e.g., system or service configuration files, system logs and user data) to steal critical data, e.g., security-critical keys, which can impair the whole system significantly, even if key management schemes [1, 2, 3] are properly established. In addition, attackers could also upload and execute malwares for long-time control. So, checking file integrity and detecting malicious files are important for the cloud-assisted CIoT system. Traditional filesystem monitors leverage agents running in operating system (OS) to perform file scanning. Since these agents are usually user space processes or kernel modules, in-VM agents could be detected or even be subverted by attackers.

Virtualization gives people a novel approach to reduce these risks. In the virtualization architecture, the virtual machine monitor (VMM) provides virtualization service for guest VMs and can access the hardware state of them. Since the VMM has the highest privilege and isolates different VMs, agentless monitors running in the VMM or a secure VM are more secure and transparent. This monitor approach is called as virtual machine introspection (VMI) [4]. Because of the security and transparency, agentless monitors are optimal for cloud computing.

Agent-less filesystem monitor tools can be classified into two categories: real-time monitors and periodic monitors. Real-time monitors [5, 6, 7, 8, 9] check access policy and file integrity in real time, most of which are usually achieved by hooking the guest VM system calls. Since system calls are invoked with high frequency, these real-time monitors always introduce performance overhead to monitored VMs. As a result, real-time monitors are unpractical to be used in cloud. Agent-less periodic monitor tools [10, 11] scan the filesystem periodically. They obtain the monitored files outside and then check the integrity by comparing the current content of them with previously gathered. To detect malwares, they calculate the file's signature and check it based on signature database. Although these approaches cannot protect files and detect malwares in real time, they introduce very low overhead to the monitored VMs. Therefore, agentless periodic monitor tools are widely used in the cloud [12, 13]. However, since all of the monitored files are scanned in every poll cycle, these monitors check lots of secure files again and again, which is needless.

To meet the filesystem monitoring requirements of cloud computing (e.g., high performance, security and low overhead), we propose a novel agentless periodic VM filesystem monitor framework, which reduces the scanning file number and improves the performance. The core idea is that our monitor only scans the modified or newly created files in every poll. So, if the secure files are not modified, they will not be scanned again and again. For protecting critical files, if the protected files are not modified, we need not to check the integrity of them. For virus scanning, since a VM has lots of system files and most of them are unmodified, checking only modified files can effectively reduce the scanning payload. As a periodic out-of-VM monitor, our system is transparent and secure.

To determine if a file needs to be scanned, we check if it has dirty blocks or clusters. If a file is modified or newly created, the disk blocks (Linux EXT4) or clusters (Windows NTFS) belong to it will be modified. In this paper, We call the dirty Linux blocks and dirty Windows clusters as dirty blocks. There are two main kinds of VM image formats: single image (RAW or QCOW2) and multiple copy-on-write images (QCOW2). Single-image VM has only one disk image, and there is no secure mechanism to log the disk modification information events. We design a backend-driver-based monitor, which hooks the backend driver to intercept the disk modification information with very low overhead, to obtain the dirty blocks. Multiple-image VM has a base image and a copy-on-write (COW) overlay. The overlay records the disk modification information after the VM initialized. By analyzing the overlay, we propose an overlay-based monitor, which can obtain the dirty blocks. Our system has high scalability supporting both Linux and Windows VMs. To enhance the system, we propose several methods to restore the modified files and remove the malicious files for cloud users.

The main contributions of our work are listed as follows:

- We propose an agentless high-performance VM filesystem monitoring method for CSPs to reduce the scanning file number in every poll. As an out-of-VM periodic tool, our monitor is transparent and secure with very low overhead.

- We design the monitor for single-image and multi-image VMs separately, which supports Linux and Windows operating systems. In addition, we propose several methods to recover the modified files and remove the malicious files.

- After implementation, the effectiveness and performance are evaluated extensively. The results show that our system can reduce the scanning file numbers significantly.

The rest of this paper is structured as follows: Section 2 summarizes the related work. The threat model and system overview is described in Section 3. Section 4 gives the single-image monitor, and the multiple-image monitor is presented in Section 5. Section 6 discusses how to handle the abnormal files. Section 7 evaluates the effectiveness and performance of our system. Conclusions and future work are given in Section 8.

## 2. Related work

Many efforts have been proposed to check file integrity because of its importance for the security of IoT ecosystems, including cloud [14], wireless sensor networks [15], mobile computing [16, 17], etc.

In general, traditional filesystem monitor runs as a kernel module or an agent in the OS. For example, XenRIM [18] runs in Xen environment where the agents run in VMs to send messages to the server running in Dom0. Unlike XenRIM, ICAR [19] runs as a kernel module to check file integrity, which can restore the original version of the modified file. Unfortunately, these in-VM approaches are not secure because they can be detected and attacked by the coexisting malwares.

With the development of virtual machine introspection (VMI) [4] technique, agentless security tools are widely be used because of the security and transparency features, such as virtual machine monitoring [20, 21], virus analysis [22], etc. In addition, CloudVMI [23] provides the VMI service for cloud users.

In general, there are two agentless filesystem monitoring ways: real-time monitoring and periodic monitoring. Real-time monitors protect files in real time. Most of them intercept file operations during the VM execution process to check rules. For instance, Flogger [5] captures VM file operations and then records the events into log files. Filesafe [6] captures file operation request to disk blocks and checks whether the request violates the policies. vMon [7] hooks QEMU [24] I/O handlers to do secure check and bridges the semantic gap between the block level and filesystem level by designing a File-to-Block Mapper. Different from these monitors, CFWatcher [8] is a target-based real-time file monitor, which only intercepts file operations of monitored files. However, these monitors are difficult to be used in practical cloud, because they introduce high overheads to the monitored VMs.

In comparison, agentless periodic monitors compare the file content with the original version periodically to detect file modifications and malicious files. CFMT [10] periodically checks the protected file integrity by comparing every protected file's checksum with the original one stored in the protected file itself. [11] checks file integrity with low security level by comparing them with secure database. Periodic monitors introduce low overheads to the monitored VMs, so they are widely used in the cloud, such as VMware vShield Endpoint [12] and Trend Micro Deep Security [13]. Since all of the monitored files are scanned in every poll cycle, the monitors will scan lots of files which does not need to scan.

Different from the existing works above, this paper aims to propose a high-performance agentless filesystem monitor to reduce the scanning file number of periodic monitors.

## 3. Threat Model and System Overview

*3.1. Assumptions and Threat Model*

In our system, we assume that the VMM is secure and can protect both itself and our system running in it from tampering by the in-VM attacks, which are able to subvert the target VM's kernel. In addition, the privileged VM (Dom0) is secure. The VMM can isolate the privileged VM from the untrusted VM and protect the privileged VM from being attacked by the untrusted VM. The communication channel between VMM and Dom0 should also be secure.

The untrusted VM is secure when it is initialized, and the initial files in it are secure. But the VM could be attacked or controlled by malwares after it is initialized. Some rootkits may subvert the VM's kernel to hide some processes and files. Therefore, the system calls related with the untrusted VM's file operations (such as OPEN, READ and WRITE) are untrusted. Furthermore, the untrusted VM's kernel memory and disk could be accessed and subverted by attackers.

*3.2. Core Idea and Challenges*

To improve the performance of periodic VM filesystem monitor, we minimize the number of scanning files of every poll cycle. Our core method is to reduce the number of scanning files by making the periodic monitor only scan the modified or newly created files. Therefore, the file scanner can scan less files in every poll, which will improve the monitor's performance significantly. To that end, the monitor should be able to identify if a file is modified or newly created. However, this task faces several challenges.

- **Effectiveness.** We need to ensure the identification results of modified or newly files are correct and usable. Most of common VM filesystems can record the file modification date into the file's meta data, which can be used to identify the modified files by checking if the modification time is after that of the last scanning. However, the meta data is not secure, because it is maintained by VM's filesystem. If we want to use this data directly, we need to make sure the data is recorded correctly and has not been modified by attackers. To that end, we need to ensure the VM's kernel and filesystem integrity and prevent file meta data from being tampered with in real time. But, it is not easy to do that. In addition. many kernel protection systems introduce extra overhead to the protected VM. So, we need to find another secure way to identify which file has been created or modified.

- **Performance.** The performance of our file monitor should be better than that of traditional file monitor which scans all of the filesystem or protected files in every poll. In addition, our monitor should not introduce much overhead to the protected running VM. The unacceptable overhead will make the system not practicable in real-world cloud computing.

- **Security and transparency.** The monitor should be prevented from being detected and attacked. As mentioned in Section 3, the VMs could be totally controlled by attackers. Our monitor should be secure and be able to work properly in this environment. In addition, the VM operating system does not need to cooperate with the file monitor.

- **Scalability.** The monitor needs to support most of VM operating systems without many modifications.

*3.3. System Overview*

To address these challenges, we propose two high-performance monitoring methods for different types of VM images, which are single image and multiple copy-on-write-based images. Most of cloud VMs use single images (RAW format). The single image stores all of the VM's filesystem. For these VMs, we design a backend-driver-based disk change tracker, which can record all of the disk changes in real time with very low overhead. By analyzing these changes, the monitor can identify the modified or newly created files (dirty files) directly. Copy-on-write-based images usually have at least two images for one VM. The base image stores the initial filesystem of VM, and the copy-on-write overlay stores all of the disk changes. In private cloud, this kind of images can instantiate lots of same VMs to build cluster rapidly. For these VMs, we propose an overlay-based analysis approach to detect the dirty files from the overlay automatically.

The system architecture is shown in Figure 1. There are three layers in our system, which are described as follows:

1. Monitor Layer - is responsible for scanning the images of different kinds of VMs outside to detect malicious files and check the integrity of protected files. By leveraging the collected disk modification information of VMs, it can perform minimized scanning.
2. Data Layer - contains the cryptographic hashes of protected files and backup copies of them. In the scanning stage, the monitor module leverages the cryptographic hashes to detect the modification of file content. The backup copies of protected files are used for recovering the modified files.
3. Utility Layer - equips the users with management tools and the warning output consoles.

The monitor module is implemented in each physical server to analyze different types of VM images, and is controlled by the controller module. To analyze the images of single-image VMs, the monitor module needs to get the disk modification information by hooking the VM backend driver. For the multiple-image VMs, the monitor module analyzes both the base image and the overlay of each VM to perform monitoring. The data layer and utility layer are implemented in the central management server for users to configure and manage the monitor. When the user configures the protected files, the system saves the hashes the files. If the files to be protected are not in the base image, the backup copies of them will be stored in the data layer for restoring files. The detection results will be sent to the users for further operations. Our system provides serval ways to delete malicious files and restore protected files.

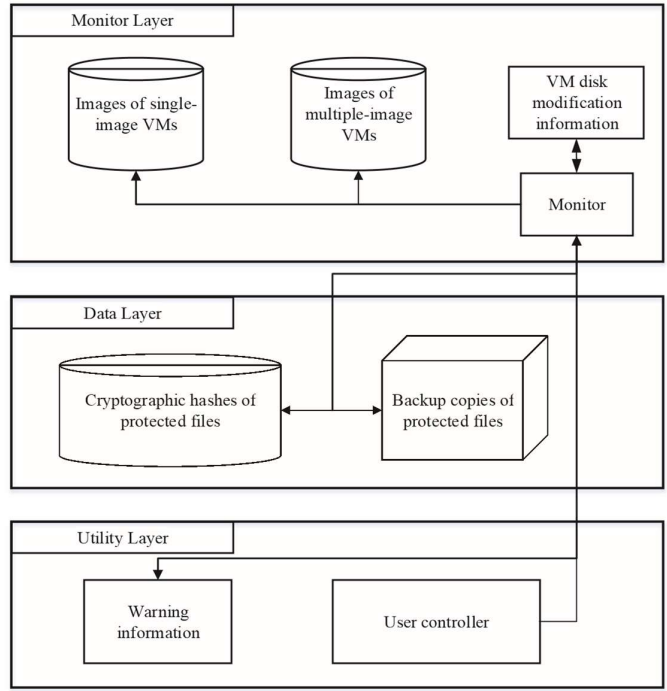

Figure 1: System overview

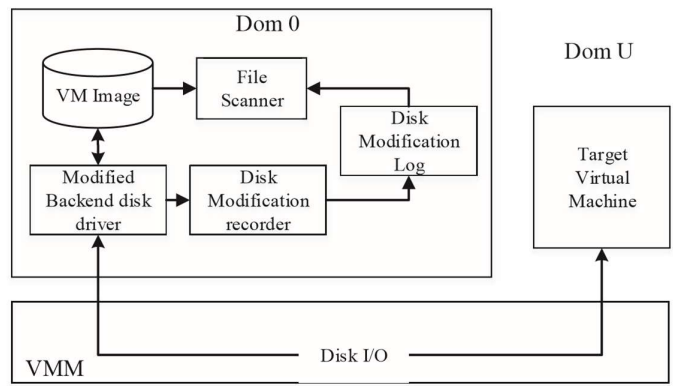

Figure 2: The architecture of monitoring single-image VMs

## 4. Monitoring Single-image Virtual Machines

*4.1. Core approach*

The format of most single-image VMs is RAW, which uses only one image to store VM filesystem. As discussed in Section 3.2, we cannot directly use the modification date recorded in the VM filesystem to identify the dirty files. To meet the secure and high-performance requirements, we design a driver-based disk modification tracker. This tracker works in the hypervisor's VM disk I/O backend driver and captures the modification information of VM disk with low overhead. Based on the disk modification information, the file scanner can identify which file is modified or newly created directly. By only checking these files, the file scanner's performance will be improved.

The system architecture is shown in Figure 2. There are three key modules in the monitor: modified backend disk driver, disk modification recorder and file scanner. The modified backend disk driver provides the disk I/O service for VMs and captures all of the modification information. The captured information will be transferred to the disk modification recorder module. The recorder leverages this information to generate the dirty block map. By analyzing this map, the file scanner can only check the modified and newly created files in every poll.

*4.2. Generating dirty block map*

To capture the dirty blocks, we modify the original VM backend disk driver to make it possible to intercept the VM disk block-level I/O operations.

The most popular two open-source virtualization hypervisors - KVM and Xen both leverage QEMU to emulate the VM disk. It supports most of mainstream formats, such as RAW and QCOW2. When a VM writes or reads a block, the request will be transferred by KVM to QEMU's related I/O driver. Since QEMU leverages files in Dom0 to emulate the VM disk, the driver converts the block-level I/O request to file I/O request. Then, it reads or writes the corresponding image file which stores the VM disk.

Our monitor hooks the writing operation in the driver to obtain two important data: offset and bytes. The offset value is the operation's start address, and the other value indicates the length of the writing content. For instance, if the offset and bytes are 1234567890 and 4096 respectively, the driver will write 4096 bytes at the address of 1234567890 of the disk image. After intercepting these two values, the driver stores them in the native pre-allocated memory. To that end, we allocate a memory area for every VM when the VM is started. When the driver is invoked to handle the writing request, it will store these two data as one record in the native memory area. The types of these two data are both unsigned long long, so one record occupies 16 bytes. The native memory space can store 100,000 records, occupying 1.6 MB, which is very small for a cloud computing server. Since the data capture is performed during the driver execution, we only record these data and do not process them to reduce the overhead.

Then, the records need to be transferred to the modification recorder. The modification recorder runs as a process in Dom0. There are several ways can be used to transfer data between processes, such as shared memory, semaphore and message queue. We first select shared memory to transfer data, because its performance is very high. The procedure is shown in Figure 3. The shared memory is established when the VM is started, and its size is 6.4 MB. The memory area is used as a circular queue and can store four entries, each of which can store 100,000 records. The shared memory can be read and written by the driver and the recorder. When the driver's native memory is full, it puts all of the records in the tail entry of the queue and then marks the entry as full. Then, it cleans the records in native memory and sends a signal to the recorder. Finally, the driver continues to handle other requests. When the recorder receives the signal, it reads all of the full entries from the queue and then marks them as empty.

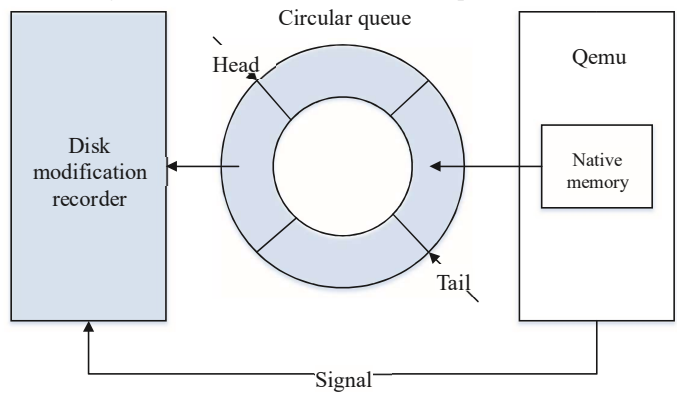

Figure 3: Using shared memory to transfer data

Besides the shared memory, we leverage files to transfer data when the shared memory is not enough to work. Since the size of the queue is fixed, if the recorder cannot clean the queue in time, the queue will be full. This case may happened when the driver handles a huge amount of requests in a very short time. When the queue is full, the driver needs to use another way to transfer data instead of waiting, because waiting will suspend the VM execution and cause high overhead. In this case, we leverage files to transfer the data. Since the size of files can be considered unlimited, all of the data can be transferred by leveraging files. When the native memory is full, the driver forks a new subprocess, which can inherit the records from the parent process, to write the data into a file. Then, the driver cleans the native memory and continues to handle other requests. The new subprocess writes 100,000 records in a file and then exits. Since the overhead of forking a new process is lower than that of writing files, the driver's performance will not be impacted significantly. In this case, the disk modification recorder needs to read files to get the records and then remove the files.

Based on the received records, the modification recorder generates a dirty block map for each VM before the file scanner starts. The dirty block map is an array, and every element of it is bool type, representing the state of a block. If a element's value is true, it indicates the corresponding block is dirty. Every block's state can be expressed as dirty block[block address] = true / false. For instance, the block (cluster) size of NTFS and EXT4 is 4096 byte. If the VM disk is configured as 10 GB, there are 2.56 M elements in the array, and the array size is 2.56 MB. Converting the records to the dirty block map is to improve the performance of file scanner. When the file scanner wants to check if a block is modified, it can use the address to check the state of it directly.

To convert the records to the dirty block map, the recorder first calculates the block address of each record by leveraging the offset obtained by the driver. There is a header at the beginning of each QEMU image, which is about 1MB (2,048 sectors). The VM disk data is stored after the header in the image. So, the block address can be calculated by (offset-header)/4096. The length of content is also important, because some operations write many blocks in a single request. For instance, if the length is 40960, ten blocks will be modified. So, if the length is not 4096, the recorder calculates the number of blocks and then marks the following several blocks as dirty blocks. The recorder processes all of the records to generate the dirty map.

*4.3. File scanner*

Our file scanner is a periodic system, which scans the VM filesystem from outside. Since all of the VM disk images are stored in Dom0 as normal files, our file scanner reads these image files in Dom0 to access the VM disks. To scan the files in the VM disk, our scanner reconstructs the filesystem of the VM and then build a file-to-block mapper for the file to be scanned. The file-to-block mapper obtains the relationship between the file and the blocks of its content. Before describe the file-to-block mapper, we first present the workflow of file scanner. The procedure is shown in Algorithm 1. File array is the list of files to be scanned. If we want to scan the whole VM filesystem, the file array will include all of the VM files. Dirty block map is generated by the disk modification recorder, which contains the state of every VM blocks. After obtaining the block addresses of the target file, the file scanner checks if the blocks contain dirty blocks. If a file has dirty blocks, it will be scanned. Otherwise, the file is secure.

**Algorithm 1** File scanning algorithm

**Input:** file array, dirty block map
**Output:** result array

```
 1: flag=False, empty(result array)
 2: for file in file array do
 3:     addr array=get content block address(file)
 4:     for addr in addr array do
 5:         if check(addr, dirty block map) == True then
 6:             flag=True
 7:             break
 8:         else
 9:             continue
10:     if flag==True then
11:         result array.add(scan(file))
12:         flag=False
13:     else
14:         result array.add(Secure)
```

15: **return** result array

---

The file-to-block mapper is responsible for getting a file's content blocks and their addresses (the third line in Algorithm 1). According to different types of VM filesystems, the file-to-block mapper works in different ways.

To build the relationship for Linux VM filesystem, we explore the Linux EXT4 filesystem. The EXT4 also called fourth extended filesystem is widely used by different Linux operating systems, such as Ubuntu and Debian. In the filesystem, every file and directory has its own data and metadata (inode), which are both recorded in the disk. To find a file in the disk, we need to follow up the front-to-back directories of the corresponding file path. Each directory has a list consisting of directory entries, associating subfile names with their inode numbers. A subdirectory is recorded in the content of its parent directory by storing the name and the corresponding inode number in its directory entry. After finding a file's inode, we can obtain the content of it by analyzing its inode. For instance, when we want to read the file "/root/test", we need first to read the directory "/" to find the subdirectory "root", and finally obtain the file "test". To read the file, we find its content blocks, which addresses are stored in the inode structure. The inode structure can store 15 block addresses, which is not enough for big files. So, these 15 blocks are classified into four types (direct blocks, indirect blocks, double indirect blocks and triple indirect blocks) to store more data. By analyzing these blocks, we can obtain a file's content block addresses.

For Windows VM filesystem, we explore the NTFS filesystem[25], which is widely used in Windows. In NTFS, all of the directories, files and their metadata are recorded as entries in the Master File Table (MFT). Each file (or directory) in MFT consists of a list of attributes. Every attribute has an attribute type, an optional attribute name, and a value. The file data is stored in the value of a certain attribute of the file. The attribute can be classified as resident attribute and non-resident attribute. A resident attribute's value is smaller than an MFT record in size, and the value will be stored within the attribute itself. In comparison, the non-resident attribute's value is stored as data runs because it is larger than an MFT record. The data attribute of a file could be resident or non-resident according to its size. When the attribute is resident, the file data is stored in the MFT record. Otherwise, it will be stored in other blocks with the address recorded in the attribute's run list.

Our file-to-block mapper can achieve the reconstruction automatically with the help of libguestfs [26], which is an open-source tool for accessing and analyzing VM disks.

The file scanner can be used to check VM file integrity and detect malicious files. As discussed in Section 3.1, we can consider that the unmodified files are secure, so they do not need to be scanned. For the modified and newly created files, we calculate the hashes and compare them with the secure ones. To check the security of a file with dirty blocks, we leverage an anti-virus software to help us. If a file is abnormal, we can handle it, which methods are described in Section 6.1 and 6.2.

*4.4. Performance and security analysis*

Our single-image VM filesystem scanner is effective, high-performance and secure with low overhead.

The disk modification information captured by the QEMU I/O backend driver is effective, because the information is collected outside and cannot be subverted. Even though the VM is controlled by attackers, the rootkits and malwares running in the target VM can only hide the files inside the VM. Our file scanner is designed to be a periodic file scanner, so we do not consider the case that the file is subverted only in VM's file cache and the modification is not written to disk. As soon as the target VM invokes a disk I/O request, our backend driver will capture it. Since our monitor is agentless and works in the secure VM, the security of it can be guaranteed by the VMM.

The file scanning process is high-performance. As described in Section 4.3, the file scanner reconstructs the VM filesystem and then checks if the target file to be scanned is modified or newly created before the scanning. The reconstruction and checking processes do not cause much extra overhead. For a traditional file scanner, it reconstructs the VM filesystem and then reads the contents for scanning directly. The main difference is the checking process. By using the dirty block map, the checking process is an O(1) algorithm, which does not read the file's content. So, the speed of checking is faster than that of the scanning. Since most of files in VM are unmodified, our scanner's speed is faster.

The disk modification tracking is real-time, so we analyze the overhead introduced to the target VM. The recorder and file scanner do not need to work during the VM execution, so the overhead is only introduced by the modified backend driver. Our modified driver records the block information in the native memory. This function is performed by several memory writing instructions, so the extra overhead caused by them is very low. When the driver needs to transfer the records to the disk modification recorder, it copies the records to the shared memory and sends a signal. But, these two operations are performed in every one hundred thousand writing requests. So the average overhead is also very low. Even though our monitor records the modification information in real time, our monitor is not a real-time system. Some real-time file monitors (such as vMon[7]) hook the I/O driver to check the block address and the block content, but the checking operation will reduce the VM I/O performance. The real-time monitors are mainly used for access control. Compared with these monitors, our monitor introduces less overhead to the target VM.

## 5. Monitoring multiple-image virtual machines

*5.1. QCOW image format and core approach*

QCOW is an image format for VM disk used by QEMU, Xen and KVM, which stands for "QEMU Copy On Write" having three versions: qcow, qcow2 and qcow3. This format leverages the Copy-On-Write (COW) feature to reduce the space in the underlying base disk image, which gives people an ability to create a base image, and create several copy-on-write overlay images on top of the base image. When a VM wants to change files, only the overlay will be changed and the original data is kept in the base image. By leveraging the copy-on-write feature, different VMs can run on only one base image with different overlay images, as shown in Figure 4. Since this image format is designed for virtual machine disks, it can support a variety of filesystems associated with different guest operating systems. Base images and overlays are widely used to instantiate thinly-provisioned VMs rapidly and create snapshots and backups of VM disks. As shown in Figure 4, the base images are read-only, and all the modifications to disks are saved in the overlays in QCOW2 format. As a result, the overlays save all the modifications after they are used by VMs, and the size becomes larger and larger.

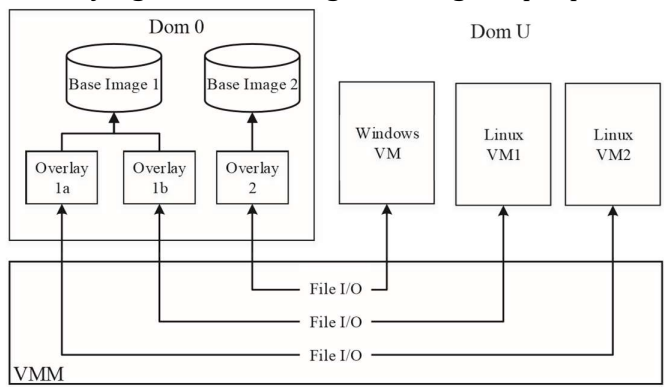

Figure 4: Relationships among base images, overlays and VMs

The overlay of the QCOW2 image format can record all of the file modifications after the VM initialization. Through analyzing the overlays, we can obtain all of the modified and created files. Since we assume that all of the files in the base image are secure, scanning only the modified and created files reduces the workload of filesystem monitor. To that end, we design the monitor to reconstruct files in the overlays. Our monitor is designed for Linux and Windows VMs separately.

*5.2. Monitoring Linux virtual machines*

For Linux, the relationship between the VM filesystem and the QCOW2 image is shown in Figure 5. In the Linux filesystem, the new file's file name and inode number are recorded in its parent directory's directory entry table. Based on the inode number and the group descriptor table, we can find the corresponding inode object from the inode table, which records the file's mode, user id and block offset, etc. To find the file content in the overlay, we need first to calculate the QCOW2 block offset based on the EXT4 filesystem block offset. In QCOW2 format, all of the data is indexed by two-level tables (e.g., L1 Table and L2 Table). Every L1 table entry points to a L2 table, and the data offset is recorded in one of the L2 table entries. The QCOW2 image data records the modifications of the disk blocks of base image. As shown in Figure 5, we can obtain the corresponding data offset by looking up the L1 and L2 tables. If the data offset is zero, the data is not allocated in the QCOW2 overlay, which means that the original file block is not modified. Otherwise, the file is modified or newly-created and is recorded in the QCOW2 image.

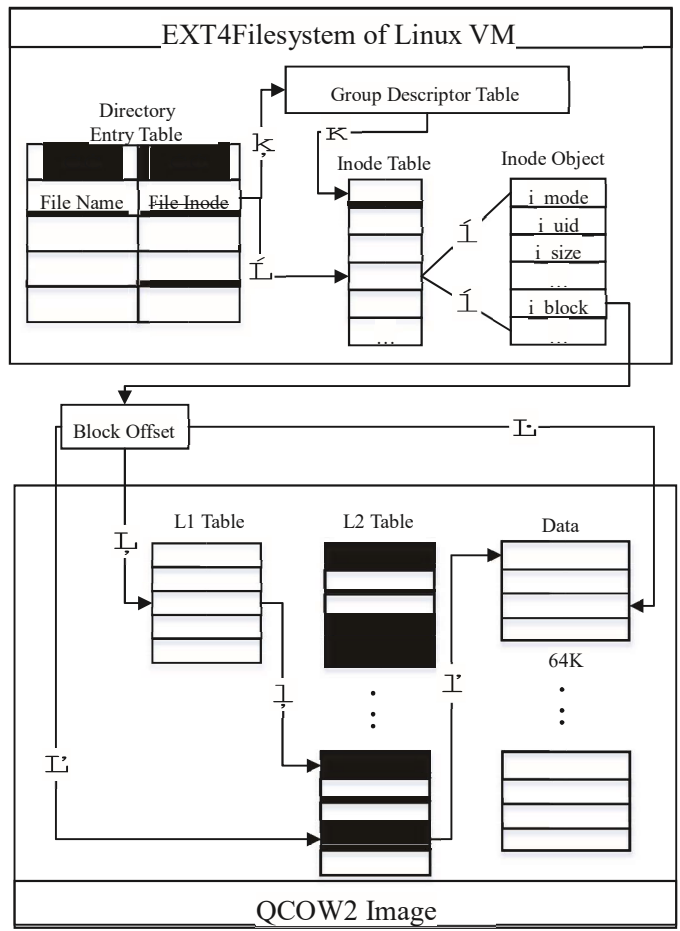

Figure 5: Relationships between Linux EXT2 filesystem and QCOW2 image

The detailed steps of scanning a file are shown in Figure 6(a). We first obtain the protected file's inode from the VM filesystem based on its file path. If the new inode cannot be found, the protected file is deleted and should be restored. Otherwise, the file's inode address can be calculated. Then, we check whether the inode object is stored in the overlay. If the inode object is not allocated in the overlay, we can ensure that the file is not modified because a modified file must have a modified inode object. However, since all of the file accesses will change the file's inode object, a modified inode object does not mean that the corresponding file content is modified. If a inode object is in the overlay but its file data is not allocated, the file is not modified. For the files with data in the overlay, we need to scan the files.

### 5.3. Monitoring Windows virtual machines

To determine whether a file is modified or newly created in Windows VM, we also search its data from the overlay. If a file's data is not recorded in the overlay, we need not to check it. The monitoring steps are similar to those of monitoring Linux VM, which are shown in Figure 6. First, we need to find the attributes of a monitored file by analyzing the VM filesystem stored in the base image and the overlay. If the file does not exist, it should be processed as a deleted file. Otherwise, we search the file's data attribute or the corresponding data blocks in the overlay according to the attribute's type. If the corresponding blocks are not allocated in the overlay, the file is not modified. Otherwise, we need to scan it.

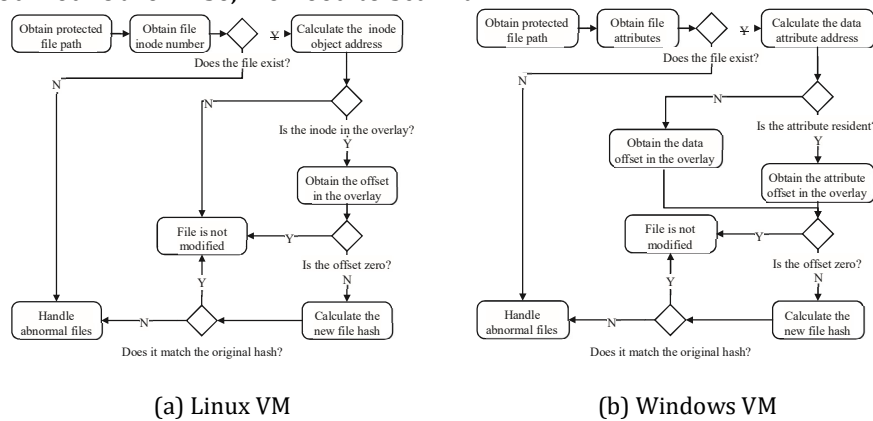

(a) Linux VM  (b) Windows VM

Figure 6: Checking protected file integrity in QCOW2 images of different VMs

The multiple-image monitor can also be used to check the file integrity and detect malicious files. The methods and steps are similar to those of the single-image scanner described in Section 4.

*5.4. Performance and security analysis*

Compared with the single-image monitor, the multiple-image monitor needs scan more files in every poll, because the overlay records all of the disk modifications after the VM initialized. Even though a new file is checked and has not been modified during the sleep time, it will be checked in next scanning. In contrast, single-image monitor will not scan this file again, because the file's blocks are not marked as dirty. However, the multipleimage monitor does not introduce any overhead to the target VM, because it does not need to trace the I/O request in real-time. As an agentless monitor, it is also very secure and transparent.

**6. Handling abnormal files**

After detecting the abnormal files, our system can recover and remove files of the protected VMs.

*6.1. Restoring protected files*

We propose a semi-automated approach to restore the modified files in an out-of-VM way. Besides alerting to the users, we package the original files in an ISO image and mount the image to the VM dynamically. Users can restore the files manually from the CD/DVD ROM of the VM. If the original files are stored in the base image, they will be fetched from the base image. Otherwise, they will be fetched from the backup copies. The files in CD/DVD ROM are non-writable, so that they cannot be tampered with. In addition, since the files to be restored can be in use by some processes, the manually restoring can ensure the VM executes properly.

*6.2. Removing malicious files*

Since malicious files may threat VM security, removing malicious files is very important for cloud users. Shutting down the VM to stop attacks is a common management approach. However, this whole-VM management approach is not optimal for some VMs, because they are providing services such as web and database services. For these VMs, the in-VM management approaches are better. Informing users to handle the malicious files is one of the in-VM management approaches. Besides, there are several agentless in-VM management approaches have been proposed, such as HyperShell [27] and EXTERIOR [28]. These approaches run in the VMM and secure VM and can manage VMs transparently, including removing files. But these management systems are complex and are difficult to be deployed in practical cloud environment.

We propose an automated real-time file removing method. Our core method is to destroy the malicious file content and then let users delete it. We destroy the file content in the VM disk and memory. We first find the file content in the disk. After that, we rewrite the blocks with zeroes. As a result, the file content is destroyed in the disk. The content of the malicious file may also be cached in the memory because the files may be in use or have been used. So, we need also destroy the file cache in the VM memory.

We leverage the VMI technique to tamper with the malicious file cache in the VM memory. We first find the data cache of malicious file in VM memory with the help of Volatility[29], which is a automated memory analysis tool. After that, we write these pages with zeroes. As a result, the malicious file's content does not exist in the VM memory. The file can be deleted by the VM users because we do not destroy the file metadata.

If the malicious files are executable malwares or malicious kernel modules, the VM may have already been attacked by attackers. To kill the malicious process or remove the rootkits, we can leverage some agentless security tools, such as [27] and [28].

**7. Evaluation**

After the implementation, we evaluate the prototype system. The tests are performed on a machine with a 2.4GHz Intel Core i5 dual-core processor supporting Intel VT with 8 GB memory. The host operating system is Ubuntu 16.04. We select an open source hypervisor - KVM with QEMU 2.8 as the testbed, which is widely used by cloud computing. To test the single-image system, we install two VMs with Ubuntu 12.04 and Windows 7 operating systems by using RAW image format. The disks are configured with 10 GB. To test the multiple-image monitor, we copy these two images as base images and create two corresponding overlays with QCOW2 image format. Then, we run another two VMs on these two overlays. All of our target VMs are configured with 1 VCPU and 1 GB memory. We will demonstrate the effectiveness and performance of our system with different VM operating systems and image formats in the following sections.

*7.1. Effectiveness*

**Case study 1: Checking file integrity**

Our scanner can be used to check file integrity. In Windows VM, we select the system critical file Host as the protected file, which is responsible for the mapping between domain names and their ip addresses. This file is often used to speed DNS resolving and block some websites. However, it is often tampered with by malwares to add phishing websites. To test the singleimage scanner, we first start the monitor, then we leverage the notpad++ to modify the Host file in the VM. Finally, we run the file scanner. The scanner finds that the HOST file has dirty blocks, which indicates that it may be modified. Actually, the HOST file is a new file. The notpad++ removes the original file and creates a new one. We do the same steps in QCOW2 VM. After the modification, we run the corresponding scanner, the result shows that the file is stored in the overlay. We can find it is modified by calculating the new hash of the file and comparing it with the one stored in the database.

In Linux VMs, we select the file "/bin/login" and a user personal file as the protected files. The "/bin/login" file is a system file and is used as a remote shell, which is usually tampered with by malwares. The user personal files including website pages are also usually tampered with by attackers. In the test, we modify these two files and then scan the filesystem. For the VM with single image, the blocks of these files are marked as dirty. For the VM with QCOW2 image, the result shows that these files are stored in the overlay.

**Case study 2: Scanning malicious files**

Our scanner can also be used to scan malicious files by using the anti-virus software to scan the newly created and modified files. In the test, we copy a malware (wcry.exe) to the Windows and Linux VMs, which is a popular malware in Windows and can be executed in Linux with wine. Our monitor detects the malware as new files in these filesystems. However, since our system can not modify the anti-virus software, we generate the modified file list and copy these file to a directory. Then, we use the anti-virus software to scan the files. The anti-virus software can mark these files as malwares.

*7.2. Performance*

Our performance test mainly focuses on the number of scanning files and scanning time of our system. The main advantage of our system is that our system can scan less files in every poll. For single-image scanner, the number of files to be scanned is related with how many files have been modified or newly created during the scanner sleeping interval. In our test, the Linux VM creates or modifies about 5 files per hour, and Windows VM has 25 dirty files per hour. The number may be very different according to different VMs with different services. But compared with the total number of VM files (more than 10,000), this number is very small. For multiple-image VMs, the number indicates how many files have been modified or newly created since the VM is initialized. The numbers of our Windows and Linux VMs are 10,072 and 5,691 respectively. These numbers may also be very different among different VMs. In addition, the number of files in overlay is always growing. Therefore, the performance of multiple-image scanner will be reduced with the execution of VM.

Detecting malicious files is a very important and common function of most anti-virus security tools. We test the performance of detecting malicious files in the whole VM filesystem with the help of ESET NOD32[30], which is a popular anti-virus tool. As mentioned in Section 7.1, our file scanner first scans the whole VM filesystem to generates the dirty file list and then uses anti-virus to scan the files. In this test, we mainly focus on multiple-image VMs, because the multiple-image VMs have more dirty files to be scanned. In Linux VM, 4.6GB of the disk is used, which has 176,647 files. Our scanner uses 217 s to find all of the files in the overlays, which number is 10,072. Then, we use NOD32 anti-virus to scan these fils, which uses 19 s. To compare with the traditional anti-virus software, we leverage NOD32 installed in the VM to rescan the filesystem in an in-VM way, and the scanning time is 10m25s. In the Windows VM test, the disk has 47,082 files occupying 6.8 GB. The scan results show that 5,691 of the files are in the overlay. The file list generating time is 134 s. Then, NOD32 uses 15 s to scan these files. To compare with the in-VM approach, we also leverage NOD32 to scan the VM filesystem. The in-VM scanning time is 273 s, which is much longer than our system. From the experimental results, we can find that the number of files in the overlay is much smaller than that of all of the files, making our system need less time. If we can make anti-virus software scan the target file just after checking the file's status, the performance will be higher.

We compare our system with an out-of-VM commercial method - VMware vShield Endpoint Data Security [12], which is a filesystem scanner for VMware ESXi VMs. Since the vShield Endpoint only supports Windows VMs, we deploy a Windows 7 VM in the VMware ESXi environment. The VM is configured with 20 GB disk, and 13.9 GB is used. After the VM tools is installed with vShield enabled, we start the data security scanning of the VM. During the scanning, we cannot see any result from the system. After the scanning is finished, we can check the log to find the start and end time of the scanning. Then, we can calculate the scanning time. The shortest scanning time is 1h 25m, which is much longer than our system. Since we cannot schedule the scanning process and cannot obtain useful output information, it is not easy to find out why this monitor needs so much time. We guess that the monitor does not want the file scanning affects the VM execution, so it lets the VM tools work very slowly.

*7.3. Overhead*

Our single-image scanner hooks the backend driver to trace the dirty blocks in real-time, which will impact the performance of the target VM's filesystem. Therefore, we test the overhead introduced by our modified backend driver in Linux VM and Windows VM. Our test focuses on the performance of writing, because we only hook the driver of block writing.

We first test the performance without hooking the backend driver. We select IOzone [31] as the benchmark for Linux VMs, which is a popular filesystem benchmark and is widely used. The test file size is set to 512 MB, and the write speeds with different block sizes are tested. After ten times of measurements, we calculate the average speeds which are shown in Figure 7. In addition, we leverage another common benchmark, decompressing kernel source code, to test the performance, because the decompression writes lots of files with different sizes. In the test, we use time command to test the execution time for ten times, the average time is 25.7s. Then, we use the CrystalDiskMark [32] as the benchmark in Windows VM, which is also a popular filesystem benchmark. The benchmark automatically measures the filesystem performance by using different reading and writing ways. We also calculate the average values of ten measurements, which are shown in Figure 8.

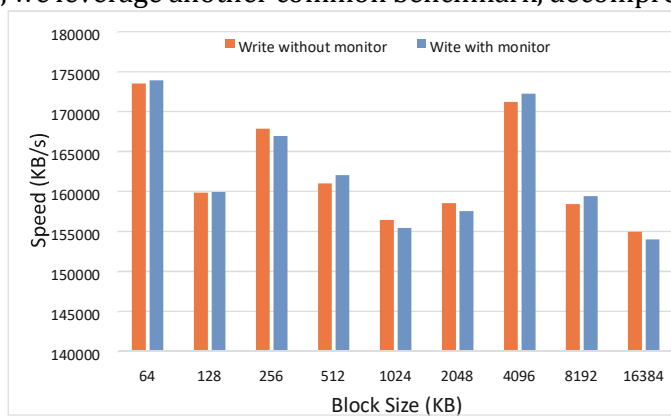

Figure 7: Overhead of Linux VM filesystem

As mentioned in Section 4.2, the backend driver intercepts the writing information and stores them in native memory, then it transfers them to the recorder. We set the file scanner's polling cycle time to one hour, which is suitable for most situations. Before scanning, the file scanner sends a signal to the recorder. The recorder outputs the dirty block map in 0.2s. This time is very short, because the recorder has already converted the records to the dirty block map and stores it in memory. After receiving the signal, it only needs to output the map to a file. Under the monitor, we leverage IOzone to test the Linux VM performance with the same measurement method, which results are shown in Figure 7. Kernel decompression time is also measured, and the average value is 26.3 s. CrystalDiskMark is used to test the Windows VM performance under the monitor. The average speeds are shown in Figure 8. From the results, we can find that the speeds with monitor are not always less than those without monitor. In some cases, the speed under monitor is even higher than the non-monitor one. The differences between speeds under monitor and speeds not under monitor are less than 3%. The results indicate that the performance degradation is very low, which is difficult to be measured.

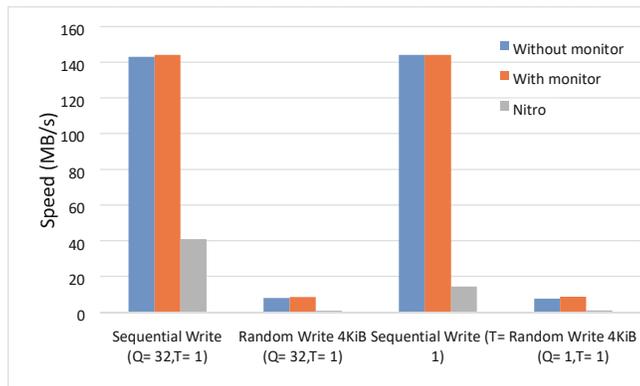

Figure 8: Overhead of Windows VM filesystem

The overhead is low because the driver only logs the information in native memory when it captures the writing request. However, if the native memory is full, the driver needs to copy the records to the shared memory and send a signal to the recorder. These operations need extra execution time. We leverage gettimeofday() to measure the execution time of these two operations for 100 times, which is about 26us in average. These operations are performed in every 100,000 writing requests. As a result, the extra overhead to every request is very low. In our test, the shared queue is long enough for the I/O requests. So the system does not use files to transfer data. But, we are also interested in its overhead. To test its overhead, we make the driver fork a new subprocess when the native memory is full. Then, we use gettimeofday() to measure the execution time of the forking operation. After 100 measurements, the average execution time is about 101us in average. The overhead is higher than that of shared memory, but it is also acceptable.

We compare our system with the system-call-based monitors, because hooking system calls is another common method to capture VM execution information. These monitors hook the VM system calls to intercept the open file list. In the test, we leverage Nitro [33] to capture the VM system calls. Nitro is a hardware-based VM system call tracker for Windows VM in KVM hypervisor. It modifies the MSR registers to make the VM trigger VMExit event when it invokes system calls. The listener running in the VMM can get the event and log the system call information. In the test, we only log the system call information of Window 7 VM and use CrystalDiskMark to test the performance. The results are show in Figure 8. From the results, we can find that the performance of system-call-based hooker is very low, that is because Nitro triggers lots of VMExits. The VMExit event suspends the VM execution, and resumes it only after the listener records all of the information. Since the VM invokes system calls frequently, the system-call-based monitor reduces the performance significantly.

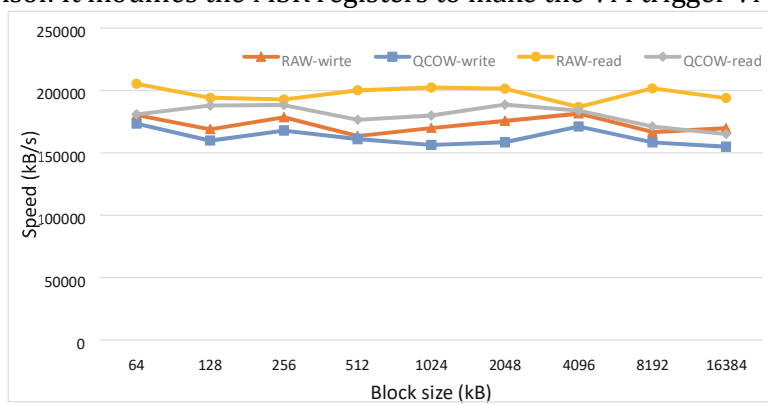

Figure 9: Linux filesystem performance comparison between RAW and QCOW2

*7.4. Discussion*

As we know, analyzing the copy-on-write image does not introduce overhead to the target VM, so we test if we can make the single-image VM turn to use copy-on-write images to reduce the monitor's overhead. To that end, we test the performance of VMs with different image formats. For Linux VM, we first run the IOzone benchmark in a Ubuntu VM with RAW image format. And then we run the benchmark in the same VM with QCOW2 enabled. The test file size is set to 512 MB, and the read/write speeds with different block sizes are shown in Figure 9. From the results, we can find that the maximal performance degradation is 15% in the write case. Then, we leverage CrystalDiskMark to test the performance difference in Window VM. The results of different disk image formats are shown in Figure 10, and the maximal performance degradation is 6.3% in the random read case.

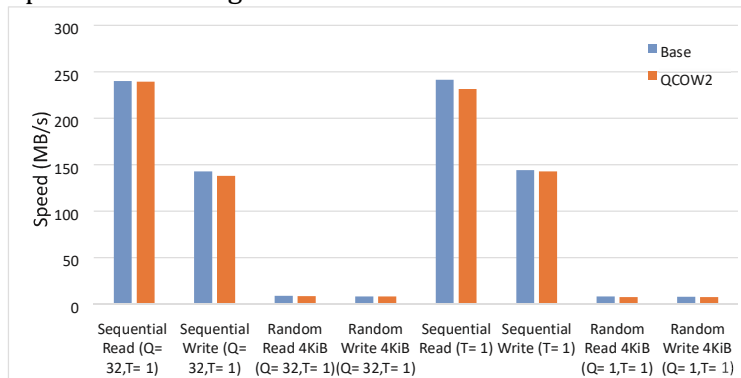

Figure 10: Windows filesystem performance comparison between RAW and QCOW2

From the comparison, the QCOW2 disk format actually causes the performance degradation, The degradation is up to 16%, which is much higher than the single-image monitor's overhead. If the VMs in cloud are using the QCOW2 images, the CSP can use the multiple-image scanner with high performance. But for the clouds using RAW images, the single-image scanner is better.

## 8. Conclusion

In this paper, we present an out-of-VM virtual machine filesystem monitor framework, which can only scan the modified and newly created files to improve the performance. We propose a backend-driver-based scanner for single-image VMs and an overlay-based scanner for multiple-image VMs. In addition, the system can handle the abnormal files, and can be used for both Linux and Windows VMs. The experimental results show that the system is effective and the scanning performance is much better than that of the traditional in-VM or out-of-VM methods. The overhead of the monitor is very low. In the future, we will continue to improve the monitor's performance.

## Acknowledgment

This work was partially supported by DongGuan Innovative Research Team Program under grants NO. 201636000100038, and National Natural Science Foundation of China under grants NO. 61601146.